\documentclass[aps,prl,twocolumn,superscriptaddress]{revtex4-1}
\usepackage{graphicx}
\usepackage{bm}
\usepackage{amsmath}
\usepackage{amssymb}
\usepackage{euscript}
\usepackage{verbatim}
\usepackage{fancyhdr}
\usepackage{wrapfig}
\usepackage{setspace}
\usepackage{xcolor}
\usepackage{amsfonts}
\usepackage{subfigure}
\usepackage{nicefrac}
\usepackage[colorlinks=true,linkcolor=blue,citecolor=blue]{hyperref}

\newcommand{\be}{\begin{equation}}
\newcommand{\ee}{\end{equation}}
\newcommand{\bea}{\begin{eqnarray}}
\newcommand{\eea}{\end{eqnarray}}

\begin{document}

\begin{titlepage}

\title{Detecting the universal fractional entropy of Majorana zero modes}

\author{Eran Sela}
    \affiliation{Raymond and Beverly Sackler School of Physics and Astronomy, Tel-Aviv University, IL-69978 Tel Aviv, Israel}
    
    \affiliation{
    	Stewart Blusson Quantum Matter Institute, University of British Columbia, Vancouver, British Columbia, Canada}
    \affiliation{
    	Department of Physics and Astronomy, University of British Columbia, Vancouver, British Columbia, Canada}

\author{Yuval Oreg}
    \affiliation{Department of Condensed Matter Physics, Weizmann Institute of Science, Rehovot, 76100, Israel}

\author{Stephan Plugge}

  \affiliation{
    	Stewart Blusson Quantum Matter Institute, University of British Columbia, Vancouver, British Columbia, Canada}
    \affiliation{
    	Department of Physics and Astronomy, University of British Columbia, Vancouver, British Columbia, Canada}

\author{Nikolaus Hartman}
    \altaffiliation{Station Q Purdue, Purdue University, West Lafayette, Indiana, USA}
    \affiliation{
    	Stewart Blusson Quantum Matter Institute, University of British Columbia, Vancouver, British Columbia, Canada}
    \affiliation{
    	Department of Physics and Astronomy, University of British Columbia, Vancouver, British Columbia, Canada}

\author{Silvia L\"uscher}
    \affiliation{
    	Stewart Blusson Quantum Matter Institute, University of British Columbia, Vancouver, British Columbia, Canada}
    \affiliation{
    	Department of Physics and Astronomy, University of British Columbia, Vancouver, British Columbia, Canada}

\author{Joshua Folk}
    \affiliation{
    	Stewart Blusson Quantum Matter Institute, University of British Columbia, Vancouver, British Columbia, Canada}
    \affiliation{
    	Department of Physics and Astronomy, University of British Columbia, Vancouver, British Columbia, Canada}

\begin{abstract}
A pair of Majorana zero modes (MZMs) constitutes a 
nonlocal qubit whose entropy is $\log 2$. Upon strongly coupling one of the constituent MZMs to a reservoir with a continuous density of states, a universal entropy change of $\frac{1}{2}\log 2$ is expected to be observed across an intermediate temperature plateau. We adapt the entropy-measurement scheme that was the basis of a recent experiment~\cite{hartman2018direct} to the case of a proximitized topological system hosting MZMs, and propose a method to measure this $\frac{1}{2}\log 2$ entropy change --- an unambiguous signature of the nonlocal nature of the topological state.  This approach offers an experimental strategy to distinguish MZMs from non-topological states.
\end{abstract}


\maketitle

\draft

\vspace{2mm}

\end{titlepage}

\emph{Introduction.--} The Majorana qubit is a nonlocal two-level system formed by two Majorana zero-modes (MZMs). These MZMs may appear, for example, in vortices of topological superconductors~\cite{read2000paired,fu2008superconducting,xu2015experimental}, as quasiparticles of exotic fractional quantum Hall states~\cite{read2000paired}, or at the edges of (quasi) 1D topological superconductors ~\cite{kitaev2001unpaired,oreg2010helical,lutchyn2010majorana,lutchyn2018majorana,mourik2012signatures,das2012zero,zhang2018quantized,hell2017two,pientka2017topological,fornieri2019evidence,ren2019topological}.
Despite an enormous body of theoretical and experimental work~\cite{alicea2012new,beenakker2013search,lutchyn2018majorana}, there is not yet conclusive evidence of the nonlocal nature of these zero modes that would distinguish them from nontopological states. In this paper, we propose an alternative direction towards this goal based on entropy measurements.

 Traditional  techniques for measuring entropy are difficult to apply to MZMs,  due to the relatively large background contribution of the phonon bath in materials or devices that would host them.  Recent progress has been achieved towards measuring the entropy of quasiparticles of exotic fractional quantum Hall states via thermalization times~\cite{schmidt2017specific}, or thermoelectric effects such as thermopower~\cite{yang2009thermopower,chickering2013thermoelectric,hou2012ettingshausen}.  Thermopower can also be used to extract entropy changes in quantum dot states~\cite{kleeorin2019measuring}. Another efficient way to measure entropy in electronic nanostructures is via the temperature dependence of charge transitions, relying on a Maxwell thermodynamic relation, $\left.{\frac{dS}{d\mu}}\right|_{T} = \left.{\frac{dN}{dT}}\right|_{\mu}$, that connects changes in the entropy, $S$, with chemical potential, $\mu$, to changes in the particle number, $N$, with temperature, $T$~\cite{cooper2009observable,ben2013detecting,kuntsevich2015strongly}. This idea was implemented in an experiment measuring  the $\log 2$ entropy of a spinful quantum dot (QD) in the Coulomb blockade regime using a charge detector~\cite{hartman2018direct}.

Here, we show theoretically that the approach in Ref.~\onlinecite{hartman2018direct} can be applied to measure the nontrivial entropy associated with MZMs at the end points of topological 1D superconductors. While our discussion focuses mainly on  semiconducting nanowires~\cite{kitaev2001unpaired,oreg2010helical,lutchyn2010majorana,lutchyn2018majorana,mourik2012signatures,das2012zero,zhang2018quantized}, the approach is general and should be operational in any system hosting MZMs, including fully-open systems like quasi-one-dimensional Josephson junctions~\cite{hell2017two,pientka2017topological,fornieri2019evidence,ren2019topological}. There are two factors that make the measurement of MZM entropy more challenging.  First, MZMs naturally come in pairs, as in the Majorana qubit, which like any two-level system has the trivial entropy $\log 2$.  Accessing the topological character of the MZM requires a measurement protocol that can resolve the entropy of an individual MZM.  We build on the well-studied problem of impurity entropy in the two channel Kondo (2CK) model~\cite{andrei1984solution,affleck1991universal}, which maps to the problem of a MZM coupled to a lead with a continuous density of states~\cite{emery1992mapping,affleck2013topological}.
In this case, a universal $\frac{1}{2} \log 2$ entropy plateau~\cite{smirnov2015majorana} can be observed, that provides the tell-tale signature of the nonlocal MZM state. 

Second, this measurement protocol is sensitive only to {\it changes} in entropy, not the absolute entropy of a state. In a spinful QD, one can start from the case of zero electrons ($N=0$, hence $S=0$) and then, using a gate voltage to add electrons, build up the entropy of higher charge states one by one. In the case of MZMs, the entropy is not directly dependent on $N$, that is, on the  parity of the MZM-hosting island.  To solve this problem, we develop a scheme in which the topological $\frac{1}{2} \log 2$ entropy of a Majorana qubit coupled to a single lead can be turned on or off by the charge on a sensor QD.  The total entropy of dot plus qubit is then $N$-dependent, providing access to the MZM state via the protocol in Ref.~\onlinecite{hartman2018direct}.

\begin{figure}[t]
	\centering
	\includegraphics[scale=0.42]{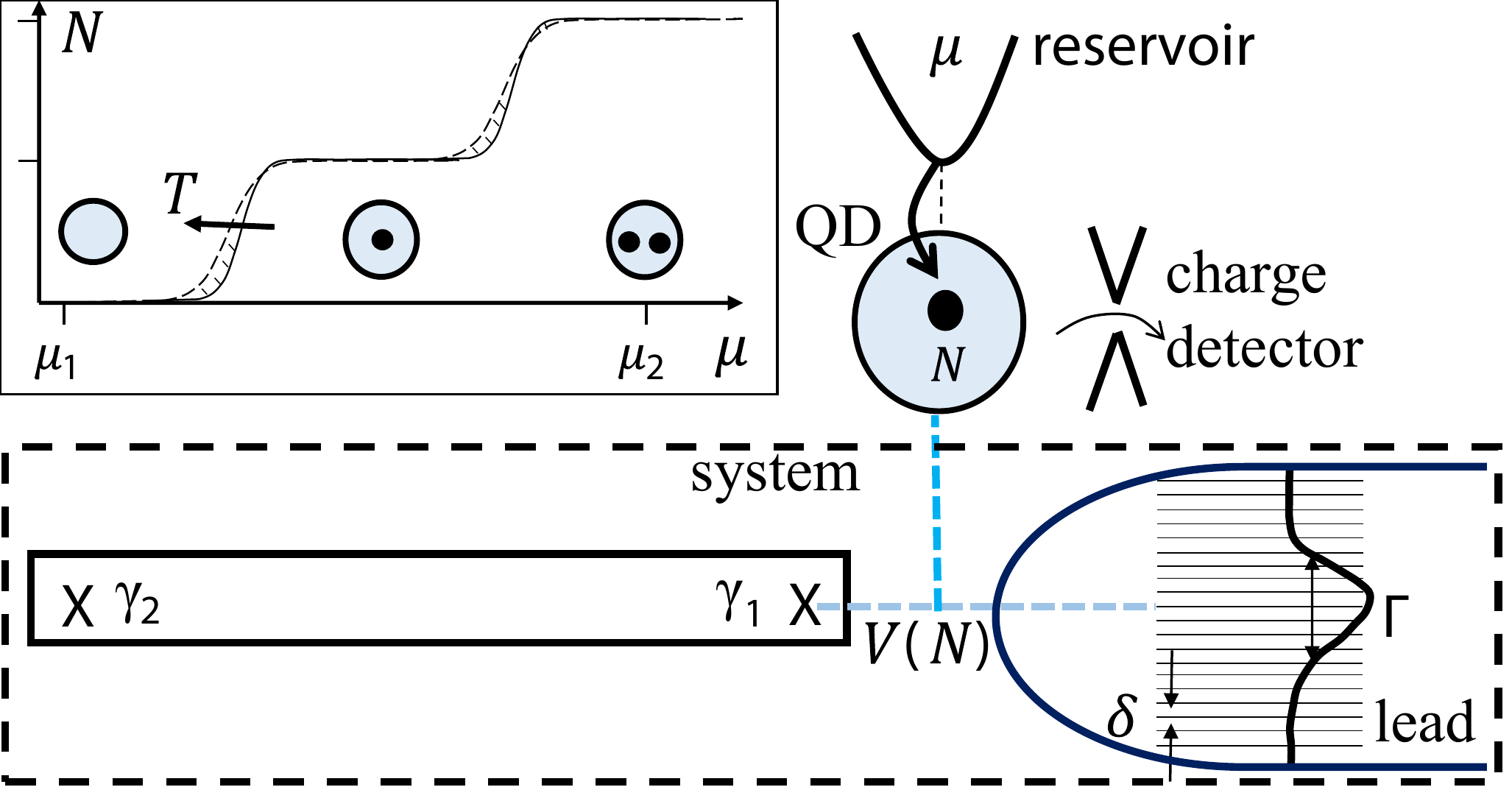}
	\caption{Schematics of the proposed entropy measurement. The ``system" consists of a Majorana qubit with two MZMs, with MZM1 coupled to a lead via coupling $V$ that is sensitive to the charge $N$ on a nearby quantum dot (QD) in the Coulomb blockade regime. QD charge steps, from $N$ to $N+1$, can be detected by a nearby sensor and induced by raising the chemical potential $\mu$ on a reservoir tunnel-coupled to the QD. In the graph, the leftward shift of the charge steps with increasing temperature indicates entropy changes [Eq.~(\ref{eq:S})].}
	\label{fig1}
\end{figure}

The measurement we propose is laid out in Fig.~1: a quantum circuit that contains a Majorana qubit (for concreteness, we consider a wire with well-separated MZMs at either end), with one end coupled to a lead across a barrier whose height depends electrostatically on the charge ($e N$) of a nearby QD.  The QD is assumed to be in the Coulomb blockade regime, described by a classical charging energy $E(N)=E_c N^2- \mu N$.  $N$ can be controlled by the chemical potential $\mu$ of a nearby reservoir from which electrons can tunnel onto the dot, although in an experiment $\mu$ would presumably be fixed and $N$ would be tuned by an electrostatic gate.  Charge steps $N \to N+1$ 
are
measured by a nearby charge detector.

In the rest of the paper, we describe how this circuit can measure the $\frac{1}{2} \log 2$ entropy of a single MZM. 
Crucially, we find that while the entropic signature of a MZM is a robust $\frac{1}{2}\log 2$, that of an Andreev bound state (ABS) accidentally tuned to zero energy may be anywhere between the trivial $\log 2$ and $\frac{1}{2}\log 2$. Low-energy ABSs are often feared to mimic MZMs in conductance measurements. A robust strategy to distinguish the two scenarios by their entropy offers an important step forward.

\emph{Entropy detection method.--}
Consider a system whose free energy, $F = F(X)$, depends on a generic parameter, $X$.
If $X$ is affected electrostatically by the QD, specifically by $N$, we have $X = X(N)$ and $F(X(N))= F(N)$.  In Fig.~1, the ``system" is delineated by the dashed box and $X$ is the coupling, $V(N)$, between $\gamma_1$ and the lead.
Within this framework, changes in the system entropy will be reflected by the temperature dependence of charge steps in the QD. While the QD affects the system electrostatically, at finite $T$ there is a thermodynamic back-action of the system on the QD: the $N$-dependent entropy of the system gives higher weight to charge states with higher entropy. 
The charge on the dot is a minimization of a thermodynamic potential, and this potential is affected both by the QD itself and by the system.


With a reservoir at chemical potential $\mu$, the total partition function of system and QD at temperature $T$ is 
\be
\label{eq:Z}
Z_{\rm{tot}}(\mu,T)=\sum_N e^{-\frac{F(N)+E(N)}{T}}.
\ee 
The QD is assumed to be spinless (that is, spin degeneracy is broken), although including QD spin would not change our results significantly.  The average number of electrons in the QD is $N(\mu)=T \frac{d\log Z_{\rm{tot}}}{d \mu}$, and the total entropy of the combined QD and system is $S_{\rm{tot}}= -dF_{\rm{tot}}/dT$, where $F_{\rm{tot}} = -T \log Z_{\rm{tot}}$. The system's entropy $S$ can be readily separated from the total by subtracting the trivial entropy of the QD, which is $\log 2$ at the charge degeneracy points and drops exponentially  to zero
away from these points.

The graph in Fig. 1 shows an example of QD charge steps, $N (\mu)$, induced by raising the reservoir chemical potential, for two different temperatures. The charge steps broaden with $T$. They also shift to the left, an effect that can be understood by integrating the Maxwell relation $\frac{dS_{\rm{tot}}}{d\mu} |_{T} = \frac{dN}{dT}|_{\mu}$ between two values of $\mu$:
\be
\label{eq:S}
\Delta S_{\rm{tot}}|_{\mu_1 \to \mu_2} = \frac{d}{d T} \int_{\mu_1}^{\mu_2} N(\mu) d \mu.
\ee
Graphically, the entropy change $\Delta S_{\rm{tot}}$ is given by the 
temperature-induced variation of the area beneath the  curve $N (\mu)$. Choosing $\mu_{1,2}$ deep inside Coulomb valleys, the horizontal leftward shift of each step with increasing temperature indicates that the system entropy is increasing with $N$.

Before proceeding with the analysis of MZM entropy detection, it is helpful to compare the experimental protocol proposed here with the measurement described in Ref.~\onlinecite{hartman2018direct}.  In that case, the measured entropy came from the spin of the QD itself; there was no external ``system" of the type shown in Fig.~1.  Ref.~\onlinecite{hartman2018direct} considered QD transitions from a spinless state with an even number $N_0$ of electrons, to a spinful state with $N_0+1$ electrons. As a result, the $N$-dependent spin degeneracy of the QD effectively makes up the system whose entropy is being measured, and at a mathematical level it can be analyzed in the same way as the present protocol. The entropic contribution to the QD charge step is thus accounted for in Eq.~(\ref{eq:Z}) by $F(N_0)=-T \log 1=0$ and $F(N_0+1)=-T \log 2$, yielding a charge step $N (\mu) =N_0+\nicefrac{2e^{-\frac{E(N_0+1)}{T}}}{\left(e^{-\frac{E(N_0)}{T}}+2e^{-\frac{E(N_0+1)}{T}}\right)}$ that shifts towards smaller $\mu$ at higher $T$. Integrating the area corresponding to this shift [Eq.~(\ref{eq:S})] gives the expected $\log 2$ entropy change as a spinful electron enters the QD \cite{hartman2018direct}.

From the point of view of the entropy measurement itself, the case of a Majorana qubit is only slightly more complicated than the simple analysis above, but from a microscopic point of view the thermodynamics of the system in Fig.~1 requires a more careful consideration.  

\emph{Entropy change of Majorana wire  side-coupled to a lead.--} Consider the total Hamiltonian $H=H_{\rm{wire}}+H_{\rm{wire-lead}}+H_{\rm{lead}}$. To describe a wire in the topological regime we consider the Kitaev chain model for $H_{\rm{wire}} $~\cite{kitaev2001unpaired,appendix}. The first site of the Kitaev chain, described by fermionic creation operator $a^\dagger_{1}$, is then coupled via normal hopping $H_{\rm{wire-lead}}=t_{\rm{wl}} a^\dagger_{1} c_1+ H.c.$ to a lead of gapless fermionic excitations, described by a half-filled tight-binding chain of length $L$, 
 $H_{\rm{lead}}=- t \sum_{j=1}^{L-1} (c^\dagger_j c_{j+1}+H.c.)$ having level spacing $\delta=\frac{2 \pi t }{L}$ for large $L$. In the analysis that follows, we report energy in units of $t$, a quarter of the bandwidth of the lead and analogous to the Fermi energy in a real system.

\begin{figure}
	\centering
	\includegraphics[width=1\linewidth]{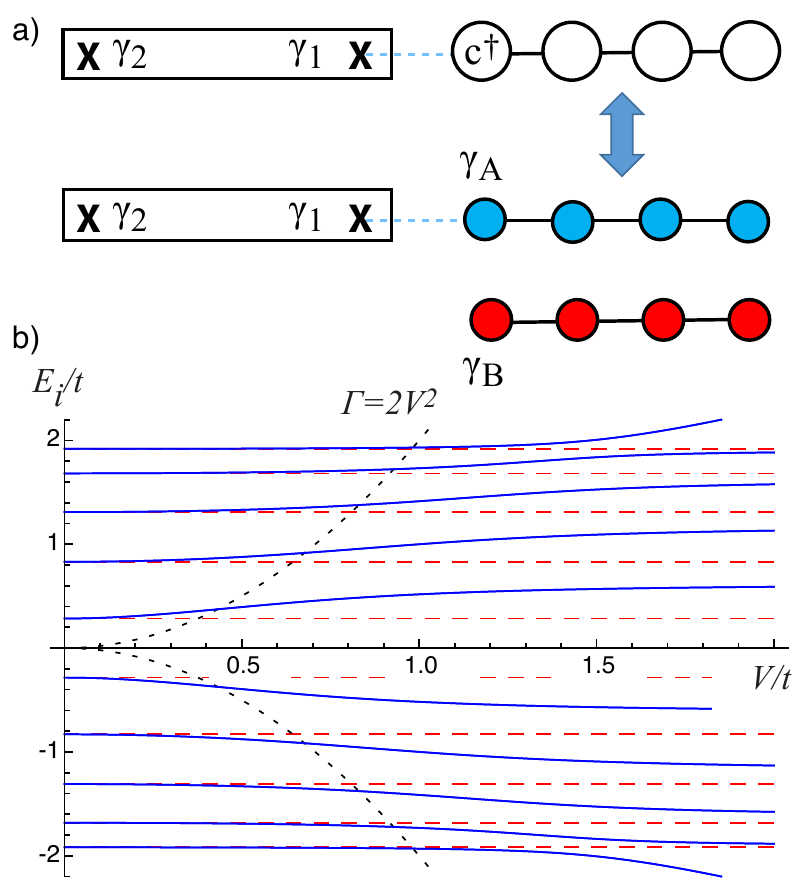}
	\caption{Absorption of a MZM into a band. (a) Effective model $H_{\rm{eff}}$ of a Kitaev chain coupled to a lead modeled by a tight-binding chain, and its equivalent in terms of Majoranas in the lead. (b) BdG spectrum for $L=10$ versus $V$. All energies are given in units of the tight binding hopping $t$.}
	\label{fig2}
\end{figure}

Within the topological regime of the wire, that is, at energy scales low compared to the energy gap of $H_{\rm{wire}}$, an effective description for the full Hamiltonian $H$ is possible in terms of the pair of MZMs $\gamma_{1,2}$ (Fig.~2a), namely, $H_{\rm{eff}} =  i V  (e^{i \phi_1}c^\dagger_1 + e^{-i \phi_1}c_1) \gamma_1+ i V_2  (e^{i \phi_2}c^\dagger_1 + e^{-i \phi_2}c_1) \gamma_2 +i \varepsilon_{12} \gamma_1 \gamma_2   + H_{\rm{lead}}$.  Here, the hopping term between $\gamma_1$ and the metallic lead is $V\propto t_{\rm{wl}}$~\cite{appendix}. The phases $\phi_1$ and $\phi_2$ are set to zero in this work, as are the couplings $\varepsilon_{12}$, between the two MZMs, and $V_2$, between  $\gamma_2$ and the lead, because both are expected to decay exponentially with the topological wire length.  As a result, we have $H_{\rm{eff}} \to  i V  (c^\dagger_1 + c_1) \gamma_1  + H_{\rm{lead}}$, unless otherwise noted.

It is instructive to first look at the evolution of the single-particle energy levels, i.e., the Bogoliubov-de Gennes (BdG) spectrum of $H_{\rm{eff}}$, as the coupling between $\gamma_1$ and the lead is turned on (Fig.~2b). For clarity, we consider the case of small $L$, where the discrete levels are clearly separated.
At $V=0$, the spectrum consists of the levels of the tight-binding chain, 
$E_j = 2 t \cos \frac{\pi j}{L+1}$ $(j=1,\dots,L)$,
and a doubly degenerate zero energy state from the decoupled MZMs.
The effect of $V$ is included by
decomposing the tight-binding chain into two Majorana chains denoted $A$ and $B$ in Fig.~2a. Without loss of generality, the latter can be defined such that $\gamma_1$ couples only to the  $A$-Majorana chain~\cite{appendix}.

When $V$ is large, the zero energy level associated with~$\gamma_1$ is absorbed into the
$A$-Majorana chain, leading to a shift of the other $A$-Majorana levels (blue in Fig.~2b) by approximately half of the level spacing in the lead. The shifting of levels  occurs up to an energy scale $\Gamma = 2V^2/t$ that depends on V, and can be interpreted as the width of $\gamma_1$.
The $B$-Majorana chain (red) is unaffected, because it decouples from $\gamma_1$ and $\gamma_2$ is not coupled to the lead.

The absorption of one MZM into the levels of the lead induces a universal change in the total entropy of the system, for temperatures greater than the level spacing in the lead but less than $\Gamma$.  This change in entropy is 
$\Delta S_V\equiv S(V) - S(V=0)$, where $S(V=0)$ is the entropy of the isolated tight-binding chain, of order $\mathcal{O}(L)$, plus an extra $\log 2$ from the decoupled MZMs. 

The drop in entropy induced by coupling to the lead, $\Delta S_V$, is plotted in Fig.~3 over a wide range of $T$ and $\Gamma$.  The curve in Fig. 3a is obtained from a numerical diagonalization of $H_{\rm{eff}}$, followed by a calculation of the entropy $S(V)=-d F/dT$ for the fermionic free energy $F = - T  \sum_{E_i  } \log (1+e^{- | E_i | /T})$~\cite{rem2}
.  This curve illustrates the characteristic signatures of MZM entropy that underpin the proposed experiment.  In the limit of low temperature ($T \ll \delta$), $\Delta S_V$ is zero because the system entropy, $S=\log 2$,  is independent of $V$: the temperature is not large enough for the chain levels to contribute, leaving only 
the pair of Majorana states at zero energy.  At temperatures larger than the level spacing but less than the width of $\gamma_1$, both $S(V)$ and $S(V=0)$ contain $\mathcal{O}(L)$ contributions from the chain levels. 
The net effect of the coupling then is a reduction by one in the number of Majorana levels within an energy window of $T$, giving $\Delta S_V=-\frac{1}{2} \log 2$ over the range $\delta \ll T \ll \Gamma$.  $\Delta S_V$ returns to zero when $T$ rises above $\Gamma$.  It is this final step in $\Delta S_V$ that is detected by the circuit in Fig.~1.

Figure 3b compares the numerical diagonalization of $H_{\rm{eff}}$ to an analytic expression for $\Delta S_V$ 
valid
in the continuum limit $\delta \ll T$ and when $\Gamma \ll t$.  Its derivation 
implies a different conceptual framework to understand the $\frac{1}{2}\log 2$ rise of $\Delta S_V$ when $\Gamma$ falls below $T$. In this approximation, the entropy change is determined by the free energy of the MZMs, $\Delta S_V = -d F_{\rm{MZM}}/dT- \log 2$, where $F_{\rm{MZM}} =-T \int_{- \infty}^\infty d E \rho(E) \log(1+e^{-|E|/T})$ is determined by the contribution of the MZMs to the density of states in the continuum limit~\cite{emery1992mapping,fabrizio1995anderson,rozhkov1997calculation,appendix}, $\rho(E) = \frac{1}{2} \delta(E)+\frac{1}{2}\frac{\Gamma /\pi}{\Gamma^2+ E^2}$.  The first term in $\rho(E)$ corresponds to the decoupled MZM, $\gamma_2$; the second  corresponds to the hybridized MZM, $\gamma_1$.  Both terms contribute $\frac{1}{2} \log 2$ to the entropy for $T \gg \Gamma$, while for $T \ll \Gamma$ only the first term contributes.

\begin{figure}[t]
	\centering
	\includegraphics[scale=1]{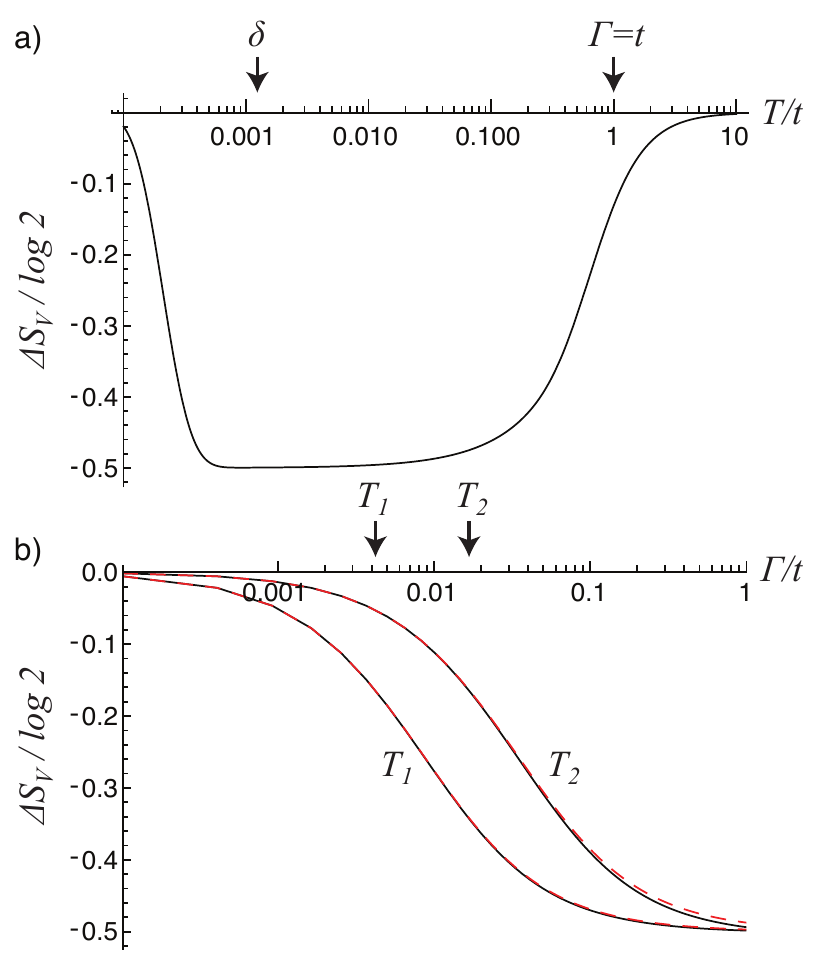}
	\caption{(a) Entropy versus temperature for the effective model $H_{\rm{eff}}$ in Fig.~2a, obtained by numerical diagonalization, illustrating the fractional $-\frac{1}{2}\log 2$ plateau at $\delta \ll T \ll \Gamma$ ($L=5000$, $t=\Gamma=1$). (b) Entropy versus wire-lead hybridization $\Gamma$ at two temperatures $T_1, T_2$. $\Gamma$ sets the width of $\gamma_1$, and as it decreases below temperature, the universal step $\frac{1}{2} \log 2$ takes place in entropy. Numerical results (solid) are compared with the analytic expression (dashed) obtained using $F_{\rm{MZM}}$ from the text ($L=1500$, $T_1=0.004 t$, $T_2=0.016 t$).}
	\label{fig3}
\end{figure}

\emph{Coulomb steps.--} The effect of the $\Gamma$-induced entropy change on the QD charge steps can be understood using Eq.~\eqref{eq:Z}, analogous to the earlier discussion for spinful QDs \cite{hartman2018direct}. For illustration, we analyze the ideal case where a single charge step in the QD results in a transition between limits $\Gamma_0 \gg T$ ($N=0$) to $\Gamma_1 \ll T$ ($N=1$). When $N=0$, $\gamma_1$ is absorbed in the lead, and the remaining free energy is $F_\mathrm{MZM}(\Gamma_0)= -\frac{T}{2}\log 2$ due to $\gamma_2$.
When $N=1$,  $F_\mathrm{MZM}(\Gamma_1)= -T\log 2$ because both MZMs are free.
Using Eq.~\eqref{eq:Z}, one finds  $N(\mu) = T\frac{d\log Z_\mathrm{tot}}{d\mu} \approx 2e^{-(E_c-\mu)/T}/(\sqrt2 + 2e^{-(E_c-\mu)/T})$ for the $N=0\to 1$ transition,
with
a charge degeneracy point $N(\mu) = \frac12$ that shifts to the left by $ -\frac{T}{2}\log 2$.
In general,
degeneracies of consecutive charge states shift by $\Delta\mu_{N,N+1} = F(N+1)-F(N) \simeq -T(S_{N+1}-S_N)$ if the main effect on free energy $F(N)$ is due to entropy $S_N$.

As a practically relevant example, a sequence of QD charge steps is simulated for a device in which $\Gamma$ depends exponentially on the barrier height,
and therefore on $N$.
Fig.~4a shows results of this simulation at two temperatures, $T_1=0.02$ and $T_2=0.04$,
where $\Gamma$ decreases from $1$ to $0.0003$ across the first two charge steps.
The entropy calculated from the integrated difference between $N(\mu)$ at the two temperatures is shown in Fig.~4b.  The entropy rise across the first peak, due to the reduction of $\Gamma$ from $1$ to $0.02$, is consistent with the shift of the charge degeneracy point (Fig.~4a inset).  The value of this entropy rise is less than the full $\frac{1}{2}\log 2$ because the crossover to $\Gamma\ll T$ is not reached until next charge step.
An additional $\log 2$ entropy at $\mu=E_c$ and $3 E_c$ is associated with QD charge degeneracy, and may be useful for calibration.

\begin{figure}[t]
	\centering
	\includegraphics[scale=0.95]{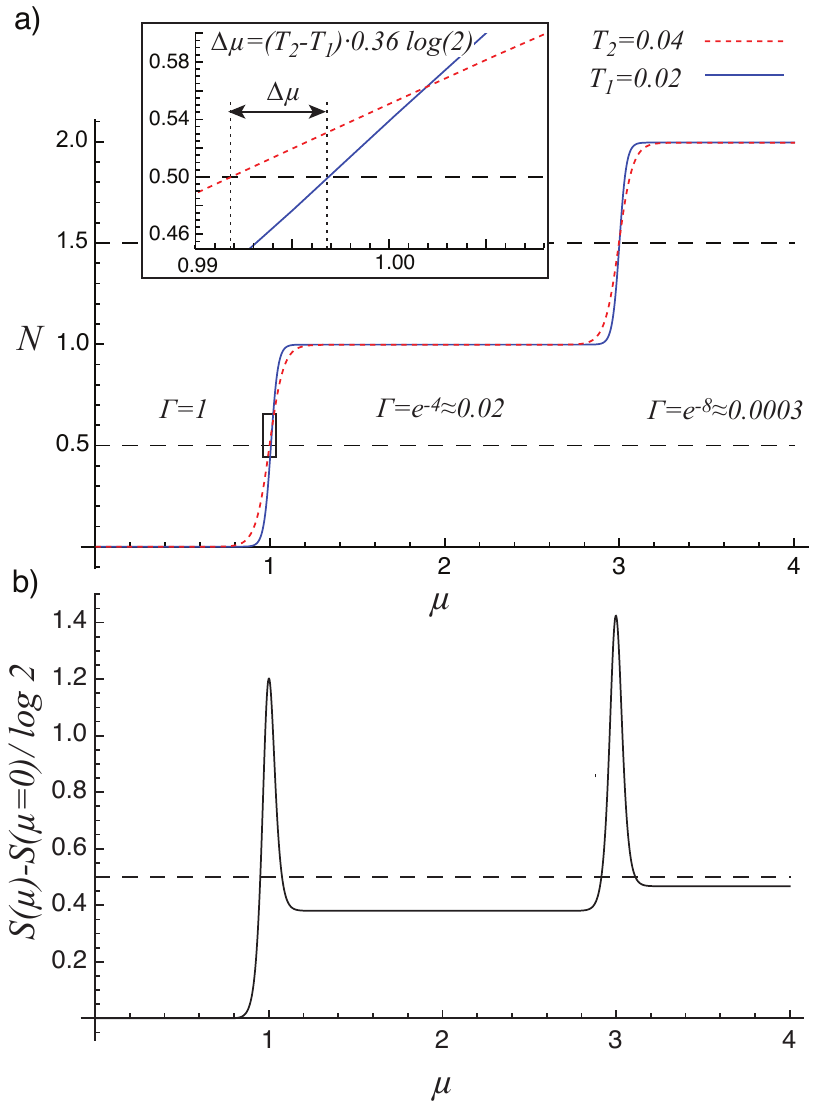}
	\caption{(a) Coulomb steps of the QD in Fig.~1 for two temperatures $T_1=0.02$ and $T_2=0.04$ ($E_c=t=1$). $\Gamma$  depends exponentially on $N$ as $\Gamma(N) = e^{-4N}$, with calculated values shown in graph. Inset: zoom-in of the first charge step, showing the shift with temperature, $\Delta \mu$, of the charge degeneracy point $N(\mu)=0.5$.
	(b) Entropy obtained by integration of $N(\mu)$'s from panel a), followed by a discrete $T-$derivative between $T_1$ and $T_2$ to approximate Eq.~(\ref{eq:S}).}
	\label{fig4}
\end{figure}

\emph{Andreev bound states.--} The entropy signature obtained for MZMs is readily distinguished from that of a regular fermionic level tuned to zero energy. Let us assume the existence of such a level, created by $d^\dagger$ and coupled to the lead via $V d^\dagger c_1 + H.c.$.
As $V$ increases, such that $\Gamma=2V^2/t$ increases from $\Gamma \ll T$ to $\Gamma \gg T$, we would find for a fermionic level tuned to zero energy that the entropy change $\Delta S_V$ for $T \gg \delta$ is 
doubled, $\Delta S_V=\log 2$~\cite{appendix}. 

This can be understood from the viewpoint of nontopological states (ABSs) as two spatially overlapping MZMs, which would generically couple to the lead with similar magnitudes~\cite{appendix}. Tuning the state to zero energy corresponds to tuning the matrix element between the two MZM wavefunctions to zero.  Depending on the non-universal ratio between the two MZM-lead couplings, 
$\Delta S_V$
could range between $\frac{1}{2} \log 2$ and $\log 2$. In contrast, a topological MZM leads to a robust entropy change of $\frac{1}{2} \log 2$ with exponentially small corrections.  Only in the case of a simultaneous coincidence \cite{appendix} -- the ABS fine-tuned to zero energy \emph{and} a particular type of asymmetric coupling between the two MZMs to the metallic lead -- does the entropic signature fail to identify the nontopological character of the ABS.

\emph{Experimental observability.--} One requirement for observing a fractionally quantized entropy change is that the temperature be in the range $\delta \ll T \ll \Gamma$. Recent measurements of quantized Majorana conductance~\cite{zhang2018quantized} imply a Majorana width $\Gamma \sim 50-100 \mu eV$; since  a metallic lead has effectively vanishing level spacing, this condition can be readily satisfied.

The second key requirement is a sensitive dependence of the wire-lead coupling on the QD charge. To achieve this, one could implement the wire-lead barrier using an unoccupied dot with virtual transport through the first level~\cite{deng2016majorana,deng2018nonlocality}.  This process is very sensitive to charge on the QD. Alternatively, the barrier's height can include a capacitive term that couples it to the entropy-measuring QD; in this case multiple Coulomb steps might be required to recover the full $\frac{1}{2}\log 2$ change in entropy.

{\it Acknowledgements:} We thank I. Affleck and A. Mitchell for discussions. This work was partially supported by BSF Grant No. 2016255 (ES), by the European Union’s Horizon 2020 research and innovation programme (grant agreement LEGOTOP No 788715), the DFG (CRC/Transregio 183, EI 519/7- 1), and the Israel Science Foundation (ISF) (YO); N.H., S.L., and J.F. were supported by Microsoft, the Canada Foundation for Innovation, the National Science and Engineering Research Council, CIFAR and SBQMI.

\bibliographystyle{apsrev4-1}

\begin{thebibliography}{36}%
	\makeatletter
	\providecommand \@ifxundefined [1]{%
		\@ifx{#1\undefined}
	}%
	\providecommand \@ifnum [1]{%
		\ifnum #1\expandafter \@firstoftwo
		\else \expandafter \@secondoftwo
		\fi
	}%
	\providecommand \@ifx [1]{%
		\ifx #1\expandafter \@firstoftwo
		\else \expandafter \@secondoftwo
		\fi
	}%
	\providecommand \natexlab [1]{#1}%
	\providecommand \enquote  [1]{``#1''}%
	\providecommand \bibnamefont  [1]{#1}%
	\providecommand \bibfnamefont [1]{#1}%
	\providecommand \citenamefont [1]{#1}%
	\providecommand \href@noop [0]{\@secondoftwo}%
	\providecommand \href [0]{\begingroup \@sanitize@url \@href}%
	\providecommand \@href[1]{\@@startlink{#1}\@@href}%
	\providecommand \@@href[1]{\endgroup#1\@@endlink}%
	\providecommand \@sanitize@url [0]{\catcode `\\12\catcode `\$12\catcode
		`\&12\catcode `\#12\catcode `\^12\catcode `\_12\catcode `\%12\relax}%
	\providecommand \@@startlink[1]{}%
	\providecommand \@@endlink[0]{}%
	\providecommand \url  [0]{\begingroup\@sanitize@url \@url }%
	\providecommand \@url [1]{\endgroup\@href {#1}{\urlprefix }}%
	\providecommand \urlprefix  [0]{URL }%
	\providecommand \Eprint [0]{\href }%
	\providecommand \doibase [0]{http://dx.doi.org/}%
	\providecommand \selectlanguage [0]{\@gobble}%
	\providecommand \bibinfo  [0]{\@secondoftwo}%
	\providecommand \bibfield  [0]{\@secondoftwo}%
	\providecommand \translation [1]{[#1]}%
	\providecommand \BibitemOpen [0]{}%
	\providecommand \bibitemStop [0]{}%
	\providecommand \bibitemNoStop [0]{.\EOS\space}%
	\providecommand \EOS [0]{\spacefactor3000\relax}%
	\providecommand \BibitemShut  [1]{\csname bibitem#1\endcsname}%
	\let\auto@bib@innerbib\@empty
	\bibitem [{\citenamefont {Hartman}\ \emph {et~al.}(2018)\citenamefont
		{Hartman}, \citenamefont {Olsen}, \citenamefont {L{\"u}scher}, \citenamefont
		{Samani}, \citenamefont {Fallahi}, \citenamefont {Gardner}, \citenamefont
		{Manfra},\ and\ \citenamefont {Folk}}]{hartman2018direct}%
	\BibitemOpen
	\bibfield  {author} {\bibinfo {author} {\bibfnamefont {N.}~\bibnamefont
			{Hartman}}, \bibinfo {author} {\bibfnamefont {C.}~\bibnamefont {Olsen}},
		\bibinfo {author} {\bibfnamefont {S.}~\bibnamefont {L{\"u}scher}}, \bibinfo
		{author} {\bibfnamefont {M.}~\bibnamefont {Samani}}, \bibinfo {author}
		{\bibfnamefont {S.}~\bibnamefont {Fallahi}}, \bibinfo {author} {\bibfnamefont
			{G.~C.}\ \bibnamefont {Gardner}}, \bibinfo {author} {\bibfnamefont
			{M.}~\bibnamefont {Manfra}}, \ and\ \bibinfo {author} {\bibfnamefont
			{J.}~\bibnamefont {Folk}},\ }\href@noop {} {\bibfield  {journal} {\bibinfo
			{journal} {Nature Physics}\ }\textbf {\bibinfo {volume} {14}},\ \bibinfo
		{pages} {1083} (\bibinfo {year} {2018})}\BibitemShut {NoStop}%
	\bibitem [{\citenamefont {Read}\ and\ \citenamefont
		{Green}(2000)}]{read2000paired}%
	\BibitemOpen
	\bibfield  {author} {\bibinfo {author} {\bibfnamefont {N.}~\bibnamefont
			{Read}}\ and\ \bibinfo {author} {\bibfnamefont {D.}~\bibnamefont {Green}},\
	}\href@noop {} {\bibfield  {journal} {\bibinfo  {journal} {Physical Review
				B}\ }\textbf {\bibinfo {volume} {61}},\ \bibinfo {pages} {10267} (\bibinfo
		{year} {2000})}\BibitemShut {NoStop}%
	\bibitem [{\citenamefont {Fu}\ and\ \citenamefont
		{Kane}(2008)}]{fu2008superconducting}%
	\BibitemOpen
	\bibfield  {author} {\bibinfo {author} {\bibfnamefont {L.}~\bibnamefont
			{Fu}}\ and\ \bibinfo {author} {\bibfnamefont {C.~L.}\ \bibnamefont {Kane}},\
	}\href@noop {} {\bibfield  {journal} {\bibinfo  {journal} {Physical review
				letters}\ }\textbf {\bibinfo {volume} {100}},\ \bibinfo {pages} {096407}
		(\bibinfo {year} {2008})}\BibitemShut {NoStop}%
	\bibitem [{\citenamefont {Xu}\ \emph {et~al.}(2015)\citenamefont {Xu},
		\citenamefont {Wang}, \citenamefont {Liu}, \citenamefont {Ge}, \citenamefont
		{Yang}, \citenamefont {Liu}, \citenamefont {Xu}, \citenamefont {Guan},
		\citenamefont {Gao}, \citenamefont {Qian} \emph
		{et~al.}}]{xu2015experimental}%
	\BibitemOpen
	\bibfield  {author} {\bibinfo {author} {\bibfnamefont {J.-P.}\ \bibnamefont
			{Xu}}, \bibinfo {author} {\bibfnamefont {M.-X.}\ \bibnamefont {Wang}},
		\bibinfo {author} {\bibfnamefont {Z.~L.}\ \bibnamefont {Liu}}, \bibinfo
		{author} {\bibfnamefont {J.-F.}\ \bibnamefont {Ge}}, \bibinfo {author}
		{\bibfnamefont {X.}~\bibnamefont {Yang}}, \bibinfo {author} {\bibfnamefont
			{C.}~\bibnamefont {Liu}}, \bibinfo {author} {\bibfnamefont {Z.~A.}\
			\bibnamefont {Xu}}, \bibinfo {author} {\bibfnamefont {D.}~\bibnamefont
			{Guan}}, \bibinfo {author} {\bibfnamefont {C.~L.}\ \bibnamefont {Gao}},
		\bibinfo {author} {\bibfnamefont {D.}~\bibnamefont {Qian}},  \emph {et~al.},\
	}\href@noop {} {\bibfield  {journal} {\bibinfo  {journal} {Physical review
				letters}\ }\textbf {\bibinfo {volume} {114}},\ \bibinfo {pages} {017001}
		(\bibinfo {year} {2015})}\BibitemShut {NoStop}%
	\bibitem [{\citenamefont {Kitaev}(2001)}]{kitaev2001unpaired}%
	\BibitemOpen
	\bibfield  {author} {\bibinfo {author} {\bibfnamefont {A.~Y.}\ \bibnamefont
			{Kitaev}},\ }\href@noop {} {\bibfield  {journal} {\bibinfo  {journal}
			{Physics-Uspekhi}\ }\textbf {\bibinfo {volume} {44}},\ \bibinfo {pages} {131}
		(\bibinfo {year} {2001})}\BibitemShut {NoStop}%
	\bibitem [{\citenamefont {Oreg}\ \emph {et~al.}(2010)\citenamefont {Oreg},
		\citenamefont {Refael},\ and\ \citenamefont {von Oppen}}]{oreg2010helical}%
	\BibitemOpen
	\bibfield  {author} {\bibinfo {author} {\bibfnamefont {Y.}~\bibnamefont
			{Oreg}}, \bibinfo {author} {\bibfnamefont {G.}~\bibnamefont {Refael}}, \ and\
		\bibinfo {author} {\bibfnamefont {F.}~\bibnamefont {von Oppen}},\ }\href@noop
	{} {\bibfield  {journal} {\bibinfo  {journal} {Physical review letters}\
		}\textbf {\bibinfo {volume} {105}},\ \bibinfo {pages} {177002} (\bibinfo
		{year} {2010})}\BibitemShut {NoStop}%
	\bibitem [{\citenamefont {Lutchyn}\ \emph {et~al.}(2010)\citenamefont
		{Lutchyn}, \citenamefont {Sau},\ and\ \citenamefont
		{Sarma}}]{lutchyn2010majorana}%
	\BibitemOpen
	\bibfield  {author} {\bibinfo {author} {\bibfnamefont {R.~M.}\ \bibnamefont
			{Lutchyn}}, \bibinfo {author} {\bibfnamefont {J.~D.}\ \bibnamefont {Sau}}, \
		and\ \bibinfo {author} {\bibfnamefont {S.~D.}\ \bibnamefont {Sarma}},\
	}\href@noop {} {\bibfield  {journal} {\bibinfo  {journal} {Physical review
				letters}\ }\textbf {\bibinfo {volume} {105}},\ \bibinfo {pages} {077001}
		(\bibinfo {year} {2010})}\BibitemShut {NoStop}%
	\bibitem [{\citenamefont {Lutchyn}\ \emph {et~al.}(2018)\citenamefont
		{Lutchyn}, \citenamefont {Bakkers}, \citenamefont {Kouwenhoven},
		\citenamefont {Krogstrup}, \citenamefont {Marcus},\ and\ \citenamefont
		{Oreg}}]{lutchyn2018majorana}%
	\BibitemOpen
	\bibfield  {author} {\bibinfo {author} {\bibfnamefont {R.}~\bibnamefont
			{Lutchyn}}, \bibinfo {author} {\bibfnamefont {E.}~\bibnamefont {Bakkers}},
		\bibinfo {author} {\bibfnamefont {L.}~\bibnamefont {Kouwenhoven}}, \bibinfo
		{author} {\bibfnamefont {P.}~\bibnamefont {Krogstrup}}, \bibinfo {author}
		{\bibfnamefont {C.}~\bibnamefont {Marcus}}, \ and\ \bibinfo {author}
		{\bibfnamefont {Y.}~\bibnamefont {Oreg}},\ }\href@noop {} {\bibfield
		{journal} {\bibinfo  {journal} {Nature Reviews Materials}\ ,\ \bibinfo
			{pages} {1}} (\bibinfo {year} {2018})}\BibitemShut {NoStop}%
	\bibitem [{\citenamefont {Mourik}\ \emph {et~al.}(2012)\citenamefont {Mourik},
		\citenamefont {Zuo}, \citenamefont {Frolov}, \citenamefont {Plissard},
		\citenamefont {Bakkers},\ and\ \citenamefont
		{Kouwenhoven}}]{mourik2012signatures}%
	\BibitemOpen
	\bibfield  {author} {\bibinfo {author} {\bibfnamefont {V.}~\bibnamefont
			{Mourik}}, \bibinfo {author} {\bibfnamefont {K.}~\bibnamefont {Zuo}},
		\bibinfo {author} {\bibfnamefont {S.~M.}\ \bibnamefont {Frolov}}, \bibinfo
		{author} {\bibfnamefont {S.}~\bibnamefont {Plissard}}, \bibinfo {author}
		{\bibfnamefont {E.~P.}\ \bibnamefont {Bakkers}}, \ and\ \bibinfo {author}
		{\bibfnamefont {L.~P.}\ \bibnamefont {Kouwenhoven}},\ }\href@noop {}
	{\bibfield  {journal} {\bibinfo  {journal} {Science}\ }\textbf {\bibinfo
			{volume} {336}},\ \bibinfo {pages} {1003} (\bibinfo {year}
		{2012})}\BibitemShut {NoStop}%
	\bibitem [{\citenamefont {Das}\ \emph {et~al.}(2012)\citenamefont {Das},
		\citenamefont {Ronen}, \citenamefont {Most}, \citenamefont {Oreg},
		\citenamefont {Heiblum},\ and\ \citenamefont {Shtrikman}}]{das2012zero}%
	\BibitemOpen
	\bibfield  {author} {\bibinfo {author} {\bibfnamefont {A.}~\bibnamefont
			{Das}}, \bibinfo {author} {\bibfnamefont {Y.}~\bibnamefont {Ronen}}, \bibinfo
		{author} {\bibfnamefont {Y.}~\bibnamefont {Most}}, \bibinfo {author}
		{\bibfnamefont {Y.}~\bibnamefont {Oreg}}, \bibinfo {author} {\bibfnamefont
			{M.}~\bibnamefont {Heiblum}}, \ and\ \bibinfo {author} {\bibfnamefont
			{H.}~\bibnamefont {Shtrikman}},\ }\href@noop {} {\bibfield  {journal}
		{\bibinfo  {journal} {Nature Physics}\ }\textbf {\bibinfo {volume} {8}},\
		\bibinfo {pages} {887} (\bibinfo {year} {2012})}\BibitemShut {NoStop}%
	\bibitem [{\citenamefont {Zhang}\ \emph {et~al.}(2018)\citenamefont {Zhang},
		\citenamefont {Liu}, \citenamefont {Gazibegovic}, \citenamefont {Xu},
		\citenamefont {Logan}, \citenamefont {Wang}, \citenamefont {Van~Loo},
		\citenamefont {Bommer}, \citenamefont {De~Moor}, \citenamefont {Car} \emph
		{et~al.}}]{zhang2018quantized}%
	\BibitemOpen
	\bibfield  {author} {\bibinfo {author} {\bibfnamefont {H.}~\bibnamefont
			{Zhang}}, \bibinfo {author} {\bibfnamefont {C.-X.}\ \bibnamefont {Liu}},
		\bibinfo {author} {\bibfnamefont {S.}~\bibnamefont {Gazibegovic}}, \bibinfo
		{author} {\bibfnamefont {D.}~\bibnamefont {Xu}}, \bibinfo {author}
		{\bibfnamefont {J.~A.}\ \bibnamefont {Logan}}, \bibinfo {author}
		{\bibfnamefont {G.}~\bibnamefont {Wang}}, \bibinfo {author} {\bibfnamefont
			{N.}~\bibnamefont {Van~Loo}}, \bibinfo {author} {\bibfnamefont {J.~D.}\
			\bibnamefont {Bommer}}, \bibinfo {author} {\bibfnamefont {M.~W.}\
			\bibnamefont {De~Moor}}, \bibinfo {author} {\bibfnamefont {D.}~\bibnamefont
			{Car}},  \emph {et~al.},\ }\href@noop {} {\bibfield  {journal} {\bibinfo
			{journal} {Nature}\ }\textbf {\bibinfo {volume} {556}},\ \bibinfo {pages}
		{74} (\bibinfo {year} {2018})}\BibitemShut {NoStop}%
	\bibitem [{\citenamefont {Hell}\ \emph {et~al.}(2017)\citenamefont {Hell},
		\citenamefont {Leijnse},\ and\ \citenamefont {Flensberg}}]{hell2017two}%
	\BibitemOpen
	\bibfield  {author} {\bibinfo {author} {\bibfnamefont {M.}~\bibnamefont
			{Hell}}, \bibinfo {author} {\bibfnamefont {M.}~\bibnamefont {Leijnse}}, \
		and\ \bibinfo {author} {\bibfnamefont {K.}~\bibnamefont {Flensberg}},\
	}\href@noop {} {\bibfield  {journal} {\bibinfo  {journal} {Physical review
				letters}\ }\textbf {\bibinfo {volume} {118}},\ \bibinfo {pages} {107701}
		(\bibinfo {year} {2017})}\BibitemShut {NoStop}%
	\bibitem [{\citenamefont {Pientka}\ \emph {et~al.}(2017)\citenamefont
		{Pientka}, \citenamefont {Keselman}, \citenamefont {Berg}, \citenamefont
		{Yacoby}, \citenamefont {Stern},\ and\ \citenamefont
		{Halperin}}]{pientka2017topological}%
	\BibitemOpen
	\bibfield  {author} {\bibinfo {author} {\bibfnamefont {F.}~\bibnamefont
			{Pientka}}, \bibinfo {author} {\bibfnamefont {A.}~\bibnamefont {Keselman}},
		\bibinfo {author} {\bibfnamefont {E.}~\bibnamefont {Berg}}, \bibinfo {author}
		{\bibfnamefont {A.}~\bibnamefont {Yacoby}}, \bibinfo {author} {\bibfnamefont
			{A.}~\bibnamefont {Stern}}, \ and\ \bibinfo {author} {\bibfnamefont {B.~I.}\
			\bibnamefont {Halperin}},\ }\href@noop {} {\bibfield  {journal} {\bibinfo
			{journal} {Physical Review X}\ }\textbf {\bibinfo {volume} {7}},\ \bibinfo
		{pages} {021032} (\bibinfo {year} {2017})}\BibitemShut {NoStop}%
	\bibitem [{\citenamefont {Fornieri}\ \emph {et~al.}(2019)\citenamefont
		{Fornieri}, \citenamefont {Whiticar}, \citenamefont {Setiawan}, \citenamefont
		{Portol{\'e}s}, \citenamefont {Drachmann}, \citenamefont {Keselman},
		\citenamefont {Gronin}, \citenamefont {Thomas}, \citenamefont {Wang},
		\citenamefont {Kallaher} \emph {et~al.}}]{fornieri2019evidence}%
	\BibitemOpen
	\bibfield  {author} {\bibinfo {author} {\bibfnamefont {A.}~\bibnamefont
			{Fornieri}}, \bibinfo {author} {\bibfnamefont {A.~M.}\ \bibnamefont
			{Whiticar}}, \bibinfo {author} {\bibfnamefont {F.}~\bibnamefont {Setiawan}},
		\bibinfo {author} {\bibfnamefont {E.}~\bibnamefont {Portol{\'e}s}}, \bibinfo
		{author} {\bibfnamefont {A.~C.}\ \bibnamefont {Drachmann}}, \bibinfo {author}
		{\bibfnamefont {A.}~\bibnamefont {Keselman}}, \bibinfo {author}
		{\bibfnamefont {S.}~\bibnamefont {Gronin}}, \bibinfo {author} {\bibfnamefont
			{C.}~\bibnamefont {Thomas}}, \bibinfo {author} {\bibfnamefont
			{T.}~\bibnamefont {Wang}}, \bibinfo {author} {\bibfnamefont {R.}~\bibnamefont
			{Kallaher}},  \emph {et~al.},\ }\href@noop {} {\bibfield  {journal} {\bibinfo
			{journal} {Nature}\ }\textbf {\bibinfo {volume} {569}},\ \bibinfo {pages}
		{89} (\bibinfo {year} {2019})}\BibitemShut {NoStop}%
	\bibitem [{\citenamefont {Ren}\ \emph {et~al.}(2019)\citenamefont {Ren},
		\citenamefont {Pientka}, \citenamefont {Hart}, \citenamefont {Pierce},
		\citenamefont {Kosowsky}, \citenamefont {Lunczer}, \citenamefont {Schlereth},
		\citenamefont {Scharf}, \citenamefont {Hankiewicz}, \citenamefont {Molenkamp}
		\emph {et~al.}}]{ren2019topological}%
	\BibitemOpen
	\bibfield  {author} {\bibinfo {author} {\bibfnamefont {H.}~\bibnamefont
			{Ren}}, \bibinfo {author} {\bibfnamefont {F.}~\bibnamefont {Pientka}},
		\bibinfo {author} {\bibfnamefont {S.}~\bibnamefont {Hart}}, \bibinfo {author}
		{\bibfnamefont {A.~T.}\ \bibnamefont {Pierce}}, \bibinfo {author}
		{\bibfnamefont {M.}~\bibnamefont {Kosowsky}}, \bibinfo {author}
		{\bibfnamefont {L.}~\bibnamefont {Lunczer}}, \bibinfo {author} {\bibfnamefont
			{R.}~\bibnamefont {Schlereth}}, \bibinfo {author} {\bibfnamefont
			{B.}~\bibnamefont {Scharf}}, \bibinfo {author} {\bibfnamefont {E.~M.}\
			\bibnamefont {Hankiewicz}}, \bibinfo {author} {\bibfnamefont {L.~W.}\
			\bibnamefont {Molenkamp}},  \emph {et~al.},\ }\href@noop {} {\bibfield
		{journal} {\bibinfo  {journal} {Nature}\ }\textbf {\bibinfo {volume} {569}},\
		\bibinfo {pages} {93} (\bibinfo {year} {2019})}\BibitemShut {NoStop}%
	\bibitem [{\citenamefont {Alicea}(2012)}]{alicea2012new}%
	\BibitemOpen
	\bibfield  {author} {\bibinfo {author} {\bibfnamefont {J.}~\bibnamefont
			{Alicea}},\ }\href@noop {} {\bibfield  {journal} {\bibinfo  {journal}
			{Reports on progress in physics}\ }\textbf {\bibinfo {volume} {75}},\
		\bibinfo {pages} {076501} (\bibinfo {year} {2012})}\BibitemShut {NoStop}%
	\bibitem [{\citenamefont {Beenakker}(2013)}]{beenakker2013search}%
	\BibitemOpen
	\bibfield  {author} {\bibinfo {author} {\bibfnamefont {C.}~\bibnamefont
			{Beenakker}},\ }\href@noop {} {\bibfield  {journal} {\bibinfo  {journal}
			{Annu. Rev. Condens. Matter Phys.}\ }\textbf {\bibinfo {volume} {4}},\
		\bibinfo {pages} {113} (\bibinfo {year} {2013})}\BibitemShut {NoStop}%
	\bibitem [{\citenamefont {Schmidt}\ \emph {et~al.}(2017)\citenamefont
		{Schmidt}, \citenamefont {Bennaceur}, \citenamefont {Gaucher}, \citenamefont
		{Gervais}, \citenamefont {Pfeiffer},\ and\ \citenamefont
		{West}}]{schmidt2017specific}%
	\BibitemOpen
	\bibfield  {author} {\bibinfo {author} {\bibfnamefont {B.}~\bibnamefont
			{Schmidt}}, \bibinfo {author} {\bibfnamefont {K.}~\bibnamefont {Bennaceur}},
		\bibinfo {author} {\bibfnamefont {S.}~\bibnamefont {Gaucher}}, \bibinfo
		{author} {\bibfnamefont {G.}~\bibnamefont {Gervais}}, \bibinfo {author}
		{\bibfnamefont {L.}~\bibnamefont {Pfeiffer}}, \ and\ \bibinfo {author}
		{\bibfnamefont {K.}~\bibnamefont {West}},\ }\href@noop {} {\bibfield
		{journal} {\bibinfo  {journal} {Physical Review B}\ }\textbf {\bibinfo
			{volume} {95}},\ \bibinfo {pages} {201306} (\bibinfo {year}
		{2017})}\BibitemShut {NoStop}%
	\bibitem [{\citenamefont {Yang}\ and\ \citenamefont
		{Halperin}(2009)}]{yang2009thermopower}%
	\BibitemOpen
	\bibfield  {author} {\bibinfo {author} {\bibfnamefont {K.}~\bibnamefont
			{Yang}}\ and\ \bibinfo {author} {\bibfnamefont {B.~I.}\ \bibnamefont
			{Halperin}},\ }\href@noop {} {\bibfield  {journal} {\bibinfo  {journal}
			{Physical Review B}\ }\textbf {\bibinfo {volume} {79}},\ \bibinfo {pages}
		{115317} (\bibinfo {year} {2009})}\BibitemShut {NoStop}%
	\bibitem [{\citenamefont {Chickering}\ \emph {et~al.}(2013)\citenamefont
		{Chickering}, \citenamefont {Eisenstein}, \citenamefont {Pfeiffer},\ and\
		\citenamefont {West}}]{chickering2013thermoelectric}%
	\BibitemOpen
	\bibfield  {author} {\bibinfo {author} {\bibfnamefont {W.}~\bibnamefont
			{Chickering}}, \bibinfo {author} {\bibfnamefont {J.}~\bibnamefont
			{Eisenstein}}, \bibinfo {author} {\bibfnamefont {L.}~\bibnamefont
			{Pfeiffer}}, \ and\ \bibinfo {author} {\bibfnamefont {K.}~\bibnamefont
			{West}},\ }\href@noop {} {\bibfield  {journal} {\bibinfo  {journal} {Physical
				Review B}\ }\textbf {\bibinfo {volume} {87}},\ \bibinfo {pages} {075302}
		(\bibinfo {year} {2013})}\BibitemShut {NoStop}%
	\bibitem [{\citenamefont {Hou}\ \emph {et~al.}(2012)\citenamefont {Hou},
		\citenamefont {Shtengel}, \citenamefont {Refael},\ and\ \citenamefont
		{Goldbart}}]{hou2012ettingshausen}%
	\BibitemOpen
	\bibfield  {author} {\bibinfo {author} {\bibfnamefont {C.-Y.}\ \bibnamefont
			{Hou}}, \bibinfo {author} {\bibfnamefont {K.}~\bibnamefont {Shtengel}},
		\bibinfo {author} {\bibfnamefont {G.}~\bibnamefont {Refael}}, \ and\ \bibinfo
		{author} {\bibfnamefont {P.~M.}\ \bibnamefont {Goldbart}},\ }\href@noop {}
	{\bibfield  {journal} {\bibinfo  {journal} {New Journal of Physics}\ }\textbf
		{\bibinfo {volume} {14}},\ \bibinfo {pages} {105005} (\bibinfo {year}
		{2012})}\BibitemShut {NoStop}%
	\bibitem [{\citenamefont {Kleeorin}\ \emph {et~al.}(2019)\citenamefont
		{Kleeorin}, \citenamefont {Thierschmann}, \citenamefont {Buhmann},
		\citenamefont {Georges}, \citenamefont {Molenkamp},\ and\ \citenamefont
		{Meir}}]{kleeorin2019measuring}%
	\BibitemOpen
	\bibfield  {author} {\bibinfo {author} {\bibfnamefont {Y.}~\bibnamefont
			{Kleeorin}}, \bibinfo {author} {\bibfnamefont {H.}~\bibnamefont
			{Thierschmann}}, \bibinfo {author} {\bibfnamefont {H.}~\bibnamefont
			{Buhmann}}, \bibinfo {author} {\bibfnamefont {A.}~\bibnamefont {Georges}},
		\bibinfo {author} {\bibfnamefont {L.~W.}\ \bibnamefont {Molenkamp}}, \ and\
		\bibinfo {author} {\bibfnamefont {Y.}~\bibnamefont {Meir}},\ }\href@noop {}
	{\bibfield  {journal} {\bibinfo  {journal} {arXiv preprint arXiv:1904.08948}\
		} (\bibinfo {year} {2019})}\BibitemShut {NoStop}%
	\bibitem [{\citenamefont {Cooper}\ and\ \citenamefont
		{Stern}(2009)}]{cooper2009observable}%
	\BibitemOpen
	\bibfield  {author} {\bibinfo {author} {\bibfnamefont {N.}~\bibnamefont
			{Cooper}}\ and\ \bibinfo {author} {\bibfnamefont {A.}~\bibnamefont {Stern}},\
	}\href@noop {} {\bibfield  {journal} {\bibinfo  {journal} {Physical review
				letters}\ }\textbf {\bibinfo {volume} {102}},\ \bibinfo {pages} {176807}
		(\bibinfo {year} {2009})}\BibitemShut {NoStop}%
	\bibitem [{\citenamefont {Ben-Shach}\ \emph {et~al.}(2013)\citenamefont
		{Ben-Shach}, \citenamefont {Laumann}, \citenamefont {Neder}, \citenamefont
		{Yacoby},\ and\ \citenamefont {Halperin}}]{ben2013detecting}%
	\BibitemOpen
	\bibfield  {author} {\bibinfo {author} {\bibfnamefont {G.}~\bibnamefont
			{Ben-Shach}}, \bibinfo {author} {\bibfnamefont {C.~R.}\ \bibnamefont
			{Laumann}}, \bibinfo {author} {\bibfnamefont {I.}~\bibnamefont {Neder}},
		\bibinfo {author} {\bibfnamefont {A.}~\bibnamefont {Yacoby}}, \ and\ \bibinfo
		{author} {\bibfnamefont {B.~I.}\ \bibnamefont {Halperin}},\ }\href@noop {}
	{\bibfield  {journal} {\bibinfo  {journal} {Physical review letters}\
		}\textbf {\bibinfo {volume} {110}},\ \bibinfo {pages} {106805} (\bibinfo
		{year} {2013})}\BibitemShut {NoStop}%
	\bibitem [{\citenamefont {Kuntsevich}\ \emph {et~al.}(2015)\citenamefont
		{Kuntsevich}, \citenamefont {Tupikov}, \citenamefont {Pudalov},\ and\
		\citenamefont {Burmistrov}}]{kuntsevich2015strongly}%
	\BibitemOpen
	\bibfield  {author} {\bibinfo {author} {\bibfnamefont {A.~Y.}\ \bibnamefont
			{Kuntsevich}}, \bibinfo {author} {\bibfnamefont {Y.}~\bibnamefont {Tupikov}},
		\bibinfo {author} {\bibfnamefont {V.}~\bibnamefont {Pudalov}}, \ and\
		\bibinfo {author} {\bibfnamefont {I.}~\bibnamefont {Burmistrov}},\
	}\href@noop {} {\bibfield  {journal} {\bibinfo  {journal} {Nature
				communications}\ }\textbf {\bibinfo {volume} {6}},\ \bibinfo {pages} {7298}
		(\bibinfo {year} {2015})}\BibitemShut {NoStop}%
	\bibitem [{\citenamefont {Andrei}\ and\ \citenamefont
		{Destri}(1984)}]{andrei1984solution}%
	\BibitemOpen
	\bibfield  {author} {\bibinfo {author} {\bibfnamefont {N.}~\bibnamefont
			{Andrei}}\ and\ \bibinfo {author} {\bibfnamefont {C.}~\bibnamefont
			{Destri}},\ }\href@noop {} {\bibfield  {journal} {\bibinfo  {journal}
			{Physical review letters}\ }\textbf {\bibinfo {volume} {52}},\ \bibinfo
		{pages} {364} (\bibinfo {year} {1984})}\BibitemShut {NoStop}%
	\bibitem [{\citenamefont {Affleck}\ and\ \citenamefont
		{Ludwig}(1991)}]{affleck1991universal}%
	\BibitemOpen
	\bibfield  {author} {\bibinfo {author} {\bibfnamefont {I.}~\bibnamefont
			{Affleck}}\ and\ \bibinfo {author} {\bibfnamefont {A.~W.}\ \bibnamefont
			{Ludwig}},\ }\href@noop {} {\bibfield  {journal} {\bibinfo  {journal}
			{Physical Review Letters}\ }\textbf {\bibinfo {volume} {67}},\ \bibinfo
		{pages} {161} (\bibinfo {year} {1991})}\BibitemShut {NoStop}%
	\bibitem [{\citenamefont {Emery}\ and\ \citenamefont
		{Kivelson}(1992)}]{emery1992mapping}%
	\BibitemOpen
	\bibfield  {author} {\bibinfo {author} {\bibfnamefont {V.}~\bibnamefont
			{Emery}}\ and\ \bibinfo {author} {\bibfnamefont {S.}~\bibnamefont
			{Kivelson}},\ }\href@noop {} {\bibfield  {journal} {\bibinfo  {journal}
			{Physical Review B}\ }\textbf {\bibinfo {volume} {46}},\ \bibinfo {pages}
		{10812} (\bibinfo {year} {1992})}\BibitemShut {NoStop}%
	\bibitem [{\citenamefont {Affleck}\ and\ \citenamefont
		{Giuliano}(2013)}]{affleck2013topological}%
	\BibitemOpen
	\bibfield  {author} {\bibinfo {author} {\bibfnamefont {I.}~\bibnamefont
			{Affleck}}\ and\ \bibinfo {author} {\bibfnamefont {D.}~\bibnamefont
			{Giuliano}},\ }\href@noop {} {\bibfield  {journal} {\bibinfo  {journal}
			{Journal of Statistical Mechanics: Theory and Experiment}\ }\textbf {\bibinfo
			{volume} {2013}},\ \bibinfo {pages} {P06011} (\bibinfo {year}
		{2013})}\BibitemShut {NoStop}%
	\bibitem [{\citenamefont {Smirnov}(2015)}]{smirnov2015majorana}%
	\BibitemOpen
	\bibfield  {author} {\bibinfo {author} {\bibfnamefont {S.}~\bibnamefont
			{Smirnov}},\ }\href@noop {} {\bibfield  {journal} {\bibinfo  {journal}
			{Physical Review B}\ }\textbf {\bibinfo {volume} {92}},\ \bibinfo {pages}
		{195312} (\bibinfo {year} {2015})}\BibitemShut {NoStop}%
	\bibitem [{app()}]{appendix}%
	\BibitemOpen
	\href@noop {} {}\bibinfo {note} {See supplemental material.}\BibitemShut
	{Stop}%
	\bibitem [{rem()}]{rem2}%
	\BibitemOpen
	\href@noop {} {}\bibinfo {note} {The sum runs over the non-negative half of
		the BdG spectrum.}\BibitemShut {Stop}%
	\bibitem [{\citenamefont {Fabrizio}\ \emph {et~al.}(1995)\citenamefont
		{Fabrizio}, \citenamefont {Gogolin},\ and\ \citenamefont
		{Nozi{\`e}res}}]{fabrizio1995anderson}%
	\BibitemOpen
	\bibfield  {author} {\bibinfo {author} {\bibfnamefont {M.}~\bibnamefont
			{Fabrizio}}, \bibinfo {author} {\bibfnamefont {A.~O.}\ \bibnamefont
			{Gogolin}}, \ and\ \bibinfo {author} {\bibfnamefont {P.}~\bibnamefont
			{Nozi{\`e}res}},\ }\href@noop {} {\bibfield  {journal} {\bibinfo  {journal}
			{Physical Review B}\ }\textbf {\bibinfo {volume} {51}},\ \bibinfo {pages}
		{16088} (\bibinfo {year} {1995})}\BibitemShut {NoStop}%
	\bibitem [{\citenamefont {Rozhkov}(1997)}]{rozhkov1997calculation}%
	\BibitemOpen
	\bibfield  {author} {\bibinfo {author} {\bibfnamefont {A.}~\bibnamefont
			{Rozhkov}},\ }\href@noop {} {\bibfield  {journal} {\bibinfo  {journal} {arXiv
				preprint cond-mat/9711181}\ } (\bibinfo {year} {1997})}\BibitemShut {NoStop}%
	\bibitem [{\citenamefont {Deng}\ \emph {et~al.}(2016)\citenamefont {Deng},
		\citenamefont {Vaitiek{\.e}nas}, \citenamefont {Hansen}, \citenamefont
		{Danon}, \citenamefont {Leijnse}, \citenamefont {Flensberg}, \citenamefont
		{Nyg{\aa}rd}, \citenamefont {Krogstrup},\ and\ \citenamefont
		{Marcus}}]{deng2016majorana}%
	\BibitemOpen
	\bibfield  {author} {\bibinfo {author} {\bibfnamefont {M.}~\bibnamefont
			{Deng}}, \bibinfo {author} {\bibfnamefont {S.}~\bibnamefont
			{Vaitiek{\.e}nas}}, \bibinfo {author} {\bibfnamefont {E.~B.}\ \bibnamefont
			{Hansen}}, \bibinfo {author} {\bibfnamefont {J.}~\bibnamefont {Danon}},
		\bibinfo {author} {\bibfnamefont {M.}~\bibnamefont {Leijnse}}, \bibinfo
		{author} {\bibfnamefont {K.}~\bibnamefont {Flensberg}}, \bibinfo {author}
		{\bibfnamefont {J.}~\bibnamefont {Nyg{\aa}rd}}, \bibinfo {author}
		{\bibfnamefont {P.}~\bibnamefont {Krogstrup}}, \ and\ \bibinfo {author}
		{\bibfnamefont {C.~M.}\ \bibnamefont {Marcus}},\ }\href@noop {} {\bibfield
		{journal} {\bibinfo  {journal} {Science}\ }\textbf {\bibinfo {volume}
			{354}},\ \bibinfo {pages} {1557} (\bibinfo {year} {2016})}\BibitemShut
	{NoStop}%
	\bibitem [{\citenamefont {Deng}\ \emph {et~al.}(2018)\citenamefont {Deng},
		\citenamefont {Vaitiek{\.e}nas}, \citenamefont {Prada}, \citenamefont
		{San-Jose}, \citenamefont {Nyg{\aa}rd}, \citenamefont {Krogstrup},
		\citenamefont {Aguado},\ and\ \citenamefont {Marcus}}]{deng2018nonlocality}%
	\BibitemOpen
	\bibfield  {author} {\bibinfo {author} {\bibfnamefont {M.-T.}\ \bibnamefont
			{Deng}}, \bibinfo {author} {\bibfnamefont {S.}~\bibnamefont
			{Vaitiek{\.e}nas}}, \bibinfo {author} {\bibfnamefont {E.}~\bibnamefont
			{Prada}}, \bibinfo {author} {\bibfnamefont {P.}~\bibnamefont {San-Jose}},
		\bibinfo {author} {\bibfnamefont {J.}~\bibnamefont {Nyg{\aa}rd}}, \bibinfo
		{author} {\bibfnamefont {P.}~\bibnamefont {Krogstrup}}, \bibinfo {author}
		{\bibfnamefont {R.}~\bibnamefont {Aguado}}, \ and\ \bibinfo {author}
		{\bibfnamefont {C.}~\bibnamefont {Marcus}},\ }\href@noop {} {\bibfield
		{journal} {\bibinfo  {journal} {Physical Review B}\ }\textbf {\bibinfo
			{volume} {98}},\ \bibinfo {pages} {085125} (\bibinfo {year}
		{2018})}\BibitemShut {NoStop}%
\end{thebibliography}

%

\newpage

\appendix

\section{Appendix I: Analytic form of the MZM entropy}
\label{apendix0}
For completeness, in this appendix we derive the analytic expression for the MZM's contribution to the free energy $F_{\rm{MZM}}(\Gamma)$ and entropy. 
We adopt calculations borrowed from the 2CK model~\cite{emery1992mapping,fabrizio1995anderson,rozhkov1997calculation}. 

Consider the model $H=H_{\rm{wire}}+H_{\rm{wire-lead}}+H_{\rm{lead}}$. The Kitaev chain offers a simple effective model for a topological wire, 
\be
\label{HK}
H_{\rm{wire}} =  \sum_{i=1}^{L_{\rm{SC}}-1} (-J a^\dagger_{j} a_{j+1}+ \Delta a^\dagger_{j} a^\dagger_{j+1} + H.c.)-\epsilon \sum_{i=1}^{L_{\rm{SC}}} a^\dagger_{j} a_{j},
\ee
which exhibits a topological transition 
at $\epsilon = \pm 2 J$.   
A pair of MZMs appears near the spatial edges of the chain for $|\epsilon|< 2 J$; the energy gap is given by $|\epsilon-2 J|$. The end of the chain is coupled via normal hopping 
\be
H_{\rm{wire-lead}}=t_{\rm{wl}} a^\dagger_{1} c_1+ H.c.
\ee 
to a lead of gapless fermionic excitations, described by a tight-binding chain of length $L$, 
\be
H_{\rm{lead}}=- t \sum_{j=1}^{L-1} (c^\dagger_j c_{j+1}+H.c.).
\ee

Setting parameters to the sweet spot of the Kitaev model $J = \Delta$, $\epsilon=0$, we have $H_{\rm{wire}} = - \sum_{i=1}^{L_{SC}-1} J (a^\dagger_j+a_j) (a_{j+1}-a_{j+1}^\dagger)$. In this case $\gamma_1 = \frac{a_1^\dagger-a_{1}}{\sqrt{2}i}$ and $\gamma_2 = \frac{a_{L_{SC}}^\dagger+a_{L_{SC}}}{\sqrt{2}}$ are the free MZMs at the ends of the wire. We use the convention $\{ \gamma_i , \gamma_j \} = \delta_{ij}$. 

At energies smaller than the topological gap, we project the wire-lead Hamiltonian  $H_{\rm{wire-lead}}=t_{\rm{wl}} a^\dagger_1 c_1+ H.c.$  into a coupling of MZM $\gamma_1$ to the lead, as $H_{\rm{wire-lead}} \to i \sqrt{\Gamma t}  \frac{c^\dagger_1 + c_1}{\sqrt{2}} \gamma_1 $. At the sweet spot $\sqrt{\Gamma t} =t_{\rm{wl}}$. Away from the sweet spot the wave function of the coupled MZM has a finite length  $\xi$ measured in units of the lattice constant, and in this case the effective coupling is $\sqrt{\Gamma t} \sim t_{\rm{wl}} / \sqrt{\xi}$.

Starting from the effective model 
\be
H_{\rm{eff}} =   i \sqrt{\Gamma t} \frac{c^\dagger_1 + c_1}{\sqrt{2}} \gamma_1   + H_{\rm{lead}},
\ee
(which is the same as in the main text with a notation change $\sqrt{\Gamma t / 2} \equiv V$), we project into the low energy modes of the tight-binding chain at half filling. For open boundary conditions the eigenmodes are annihilated by $c_k = \frac{\sqrt{2}}{\sqrt{L+1}} \sum_{j=1}^L \sin \frac{\pi j n_k}{L+1} c_j$, $(n_k=1, \dots, L)$ with energy $E_k = 2 \cos \frac{\pi n_k}{L+1}$. Near wave number $k_F = \pi/2$ the dispersion is linear $\epsilon_k=-2t (k-k_F)$ with level spacing $\delta=\frac{2 \pi t}{L}$. Inverting this transformation  we have $c_{j=1} = \sum_k \frac{\sqrt{2}}{\sqrt{L+1}}  \sin \frac{\pi  n_k}{L+1} c_k$. Projecting $H_{\rm{eff}}$ into the $k$-modes defined around $k_F=\pi/2$ we have
\be
\label{Heffapp}
H_{\rm{eff}}=\sum_k \epsilon_k c^\dagger_k c_k+i \sqrt{\Gamma t} \sum_k \frac{\sqrt{2}}{\sqrt{L+1}} (c_k^\dagger + c_k)  \gamma_1 .
\ee
We use this Hamiltonian to compute the MZM Green function and contribution to the density of states. Out of the two MZMs we define a Dirac fermion
\be
\gamma_1 = \frac{d^\dagger+d}{\sqrt{2}},~~~\gamma_2 = \frac{d^\dagger-d}{\sqrt{2}i},
\ee
with $d^\dagger= \frac{\gamma_1+i \gamma_2}{\sqrt{2}}$. The retarded $d-$Green function is $G_d(\omega) =-i \int_0^\infty dt e^{- \delta t} e^{i \omega t} \langle \{d(t) , d^\dagger \} \rangle$, ($\delta = 0^+$), giving us the MZM contribution to the density of states $\rho(\omega)=-\frac{1}{\pi} {\rm{Im}}G_d(\omega) $. For $V=0$, $G_d(\omega) = (\omega + i \delta)^{-1}$, and the MZM density of states is just $\rho(\omega)=\delta(\omega)$. 

For $V \ne 0$, assuming $\epsilon_{12}=0$ so that the two MZMs are decoupled, we have $G_d(\omega)=\frac{1}{2}G_{\gamma_1}(\omega)+\frac{1}{2}G_{\gamma_2}(\omega)$ where $G_{\gamma_i}(\omega) =-i \int_0^\infty dt e^{- \delta t} e^{i \omega t}  \langle \{\gamma_i(t) , \gamma_i \} \rangle$. Also, as long as $\epsilon_{12}=0$ we have $G_{\gamma_2}(\omega) =(\omega+ i \delta)^{-1}$, namely MZM $\gamma_2$ remains decoupled. The $G_{\gamma_1}$ Green function can be computed to infinite order in $V$, $G_{\gamma_1}=\frac{1}{\omega - \Sigma}$, where the self-energy $\Sigma$ is of second order in the perturbation in Eq.~(\ref{Heffapp}) 
\be
\Sigma (\omega)= (t \Gamma) \frac{2}{L} \sum_k G_k(\omega),
\ee
and where $G_k(\omega) = (\omega - \epsilon_k+ i \delta)^{-1}$ is the Green function of the $k$-states in the leads. Replacing the sum by an integral in the infinite band limit $\sum_k = \int_{-\infty}^\infty \frac{d \epsilon_k}{\delta}$ we have simply $\Sigma(\omega) = - i \Gamma$. Thus, the contribution to the density of states from the MZMs is \be
\rho(\omega) = \frac{1}{2} \delta(\omega)+\frac{1}{2} \frac{\Gamma/\pi}{\omega^2+\Gamma^2}.
\ee

Finally, we can use this result for the MZM's contribution to the density of states to obtain their contribution to the free energy and entropy. For a single particle Hamiltonian with eigenvalues $\epsilon_i$ and density of states $\rho(\omega)= \sum_i \delta(\omega - \epsilon_i)$ the free energy is $F=-T \sum_i \log (1+e^{-\epsilon_i /T}) =-T \int d \omega \rho (\omega)  \log (1+e^{-\omega /T})$. The MZM contribution to the free energy and entropy $S=-dF/dT$ is obtained by this formula with the MZM density of states $\rho$.

\section{Appendix II: Effective model for Andreev bound states}

\begin{figure}
	\centering	
	\includegraphics[width=1\linewidth]{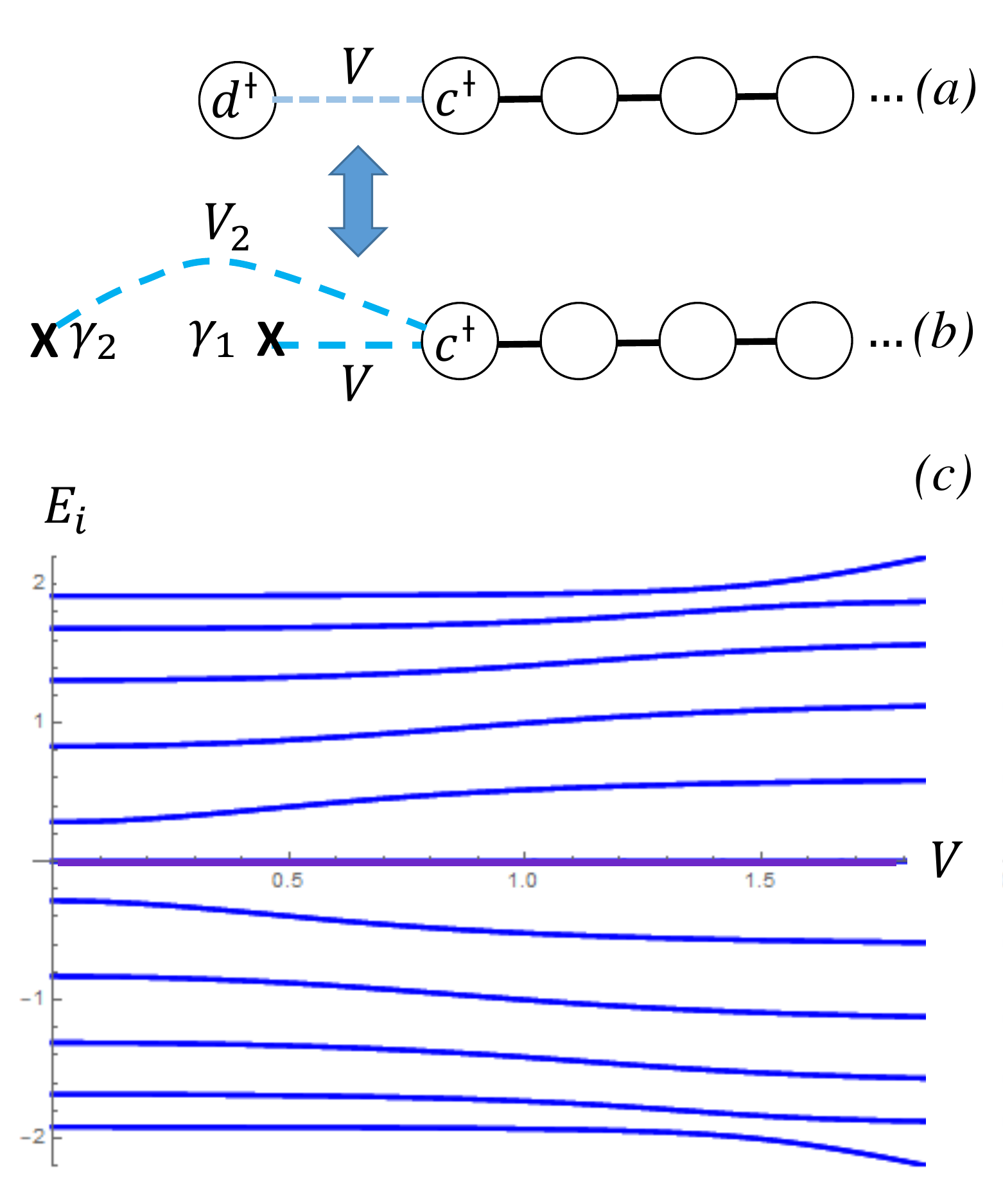}
	\caption{(a) A tight-binding chain coupled to a conventional, non-Majorana fermionic level $d^\dagger$, described by Hamiltonian $H_{\rm{eff},\rm{ABS}}$, representing a nontopological Andreev bound state.  (b) the spectrum as function of $V$ for a chain of length $L=10$. When the $d$ level gets absorbed into the band all the low energy levels move away, while in the Majorana case only half of the metallic levels (represented by $\gamma_A$ in Fig.~2a in the main text) respond.} 
	\label{fig:part2p}
\end{figure}

In order to distinguish the entropy contribution of a MZM from that of a non-topological state, we consider a regular fermionic level tuned to zero energy. It is coupled to the lead, described by Hamiltonian $H_{\rm{lead}}$ via a coupling $\delta H = V d^\dagger c_1 + H.c.$, displayed in Fig.~\ref{fig:part2p}a. Since $d$ can be decomposed into two MZMs, this model is equivalent to two MZMs coupled to the lead, see Fig.~\ref{fig:part2p}b. This is of the same form of the effective model used in the main text,
\bea
\label{eq:heff}
H_{\rm{eff}}&=&  i V  (e^{i \phi_1}c^\dagger_1 + e^{-i \phi_1}c_1) \gamma_1+ i V_2  (e^{i \phi_2}c^\dagger_1 + e^{-i \phi_2}c_1) \gamma_2   \nonumber \\
&+& H_{\rm{lead}}+i \varepsilon_{12} \gamma_1 \gamma_2.
\eea
In the extreme case of an an ABS we have $V = V_2$. Further more $\epsilon_{12}=0$ corresponds to a zero energy level. The evolution of the single-particle levels is shown in Fig.~\ref{fig:part2p}c. We can see that as $V$ increases, zero energy levels repel \emph{all} the levels of the tight binding chain, not only those of the $A$-Majorana chain as in Fig.~2b in the main text.
As a result, the plateau in $\Delta S_V$ for intermediate $T$ [c.f. Fig.~3a] is doubled, $\Delta S_V=\log 2$.

This should be contrasted with the case of a MZM. In this case $V_2=0$, and a single MZM coules to the lead. The effect of $V$ can be understood, as described in the main text, by
decomposing the tight-binding chain into two Majorana chains denoted $A$ and $B$, described by the Hamiltonian 
\bea
H_{\rm{lead}}
&=&t \sum_{j=1}^{L-1} (c^\dagger_j c_{j+1}+H.c.) \nonumber \\
&=&2it\sum_{j=1}^{L-1}( \gamma_{A j} \gamma_{A j+1} + \gamma_{B j} \gamma_{B j+1}),
\eea
with $c_j =\frac{ \gamma_{A j}+i \gamma_{B j}}{\sqrt{2}}$ for odd $j$ and $c_j =\frac{ \gamma_{B j}-i \gamma_{A j}}{\sqrt{2}}$ for even $j$. Using this decomposition of the fermionic states in the lead, we see that $\gamma_1$ couples only to the  $A$-Majorana chain (Fig.~2a in the main text), and as a result, only half of the levels move upon increasing $V$ (Fig.~2b), leading to an entropy change of $\frac{1}{2} \log 2$.

More generally, a non-topological accidental zero energy state can be represented by two MZMs which are coupled to the lead with independent coefficients $V$ and $V_2$ as in  Eq.~(\ref{eq:heff}), see Fig.~\ref{fig:part2p}b. To see this consider a superconductor on a 1D line with electronic creation operator $\psi^\dagger(x)$. The creation operator of a quasiparticle is
\be
\Gamma^\dagger= \int dx [u(x) \psi^\dagger(x)+v(x) \psi(x)].
\ee
Representing the electron creation and annihilation by Majorana fields,
\be
\psi(x)=(\alpha(x) + i \beta (x))/2,
\ee
with $\alpha(x)=\psi^\dagger(x)+\psi(x)=\alpha^\dagger(x)$ and $\beta(x)= i [\psi^\dagger(x)- \psi(x)]=\beta^\dagger(x)$   we can write $\Gamma^\dagger= \int dx [\frac{1}{2}(u(x) +v(x)) \alpha(x)+ \frac{i}{2}(v(x) -u(x)) \beta(x)] \equiv \int dx \left[\phi_\alpha(x) \alpha(x)+\phi_\beta(x) \beta (x) \right]$. A special situation is the MZM case, where the wave function $\phi_\alpha(x)$ is peaked at one end and $\phi_\beta(x)$ on the other end, so that coupling a lead to one end of the wire ensures a coupling to only one MZM.  However it is possible to have a rare situation in which $\phi_\alpha(x)$ and $\phi_\beta(x)$ overlap in space so that
$ \int \phi_\alpha(x) \phi_\beta(x) dx \ne 0$, but still the matrix element of the Hamiltonian of the wire, $H(x)$,  between these wave functions is zero
$ V_{\alpha \beta}=\int \phi_\alpha(x) H(x) \phi_\beta(x) dx =0$. In this case  one has an Andreev state with zero energy. 

An additional fine tuning may nullify, for example, the matrix element of $ \beta(x)$ with the lead. That occurs when $ \int \phi_\beta(x) V(x) \phi_{\rm lead}(x) dx=0$,
with $V(x)$ describing the tunneling potential between the lead and the wire and $ \phi_{\rm lead}(x)$ is the wave function of an electron in the lead. Only in the case of a simultaneous coincidence---the ABS fine-tuned to zero energy \emph{and} a particular type of asymmetric coupling to the metallic lead---would the entropic signature fail to identify the nontopological character of the ABS.
Essentially one then has a fine-tuned local \emph{pair} of MZMs.
\end{document}